\documentclass[lettersize,journal]{IEEEtran}
\usepackage{amsmath,amsfonts}
\usepackage{algorithmic}
\usepackage{algorithm}
\usepackage{array}
\usepackage[caption=false,font=normalsize,labelfont=sf,textfont=sf]{subfig}
\usepackage{textcomp}
\usepackage{stfloats}
\usepackage{url}
\usepackage{verbatim}
\usepackage{graphicx}
\usepackage{cite}
\usepackage{multirow}
\usepackage{booktabs}
\usepackage{xcolor}
\begin{document}

\title{MIMO Evolution toward 6G: End-User-Centric Collaborative MIMO}

\author{Lung-Sheng~Tsai,~Shang-Ling~Shih,~Pei-Kai~Liao,~\IEEEmembership{Senior Member,~IEEE},~and~Chao-Kai~Wen,~\IEEEmembership{Senior Member,~IEEE}

\thanks{{L.-S.~Tsai} and {P.-K.~Liao} are with the MediaTek Inc., Hsinchu, Taiwan, Email: {\rm Longson.Tsai@mediatek.com, pk.liao@mediatek.com}.}
\thanks{{S.-L.~Shih} and {C.-K.~Wen} are with the Institute of Communications Engineering, National Sun Yat-sen University, Kaohsiung 80424, Taiwan, Email: {\rm monlylonly@gmail.com, chaokai.wen@mail.nsysu.edu.tw}.}

}

\markboth{IEEE Communications Magazine}%
{Shell \MakeLowercase{\textit{et al.}}: A Sample Article Using IEEEtran.cls for IEEE Journals}


\maketitle

\begin{abstract}
In 6G, the trend of transitioning from massive antenna elements to even more massive ones is continued. However, installing additional antennas in the limited space of user equipment (UE) is challenging, resulting in limited capacity scaling gain for end users, despite network side support for increasing numbers of antennas. To address this issue, we propose an end-user-centric collaborative MIMO (UE-CoMIMO) framework that groups several fixed or portable devices to provide a virtual abundance of antennas. This article outlines how advanced L1 relays and conventional relays enable device collaboration to offer diversity, rank, and localization enhancements. We demonstrate through system-level simulations how the UE-CoMIMO approaches lead to significant performance gains. Lastly, we discuss necessary research efforts to make UE-CoMIMO available for 6G and future research directions.
\end{abstract}

\section{Introduction}

\IEEEPARstart{M}{assive} MIMO technology plays a crucial role in boosting spectral efficiency during the transition from 4G Long Term Evolution (LTE) to 5G New Radio (NR). Active antenna systems have made it feasible to accommodate large numbers of digital antenna ports at a base station (BS). Another approach to achieving massive MIMO is aggregating antennas from multiple transmission/reception points (multi-TRPs), which can lower implementation costs. By collaborating and controlling the multi-TRPs, a network can serve users simultaneously, achieving diversity gain or beamforming gain across multi-TRPs. Techniques such as dynamic point selection, non-coherent joint transmission, and coherent joint transmission have been realized in 5G NR Rel-16 and Rel-17 standards and further developed in Rel-18 \cite{Jin-23JSAC}.

The evolution towards more antenna elements is expected to continue in 6G, transitioning from massive to extremely massive antenna elements. To enable this transition, a flexible approach to constructing a large active antenna array utilizing standardized antenna modules in a distributed manner has been proposed \cite{Jeon-21COMMag}. However, despite the increasing number of antennas supported on the network side, there are limitations in adding more antennas on the user equipment (UE) side, such as customer premises equipments (CPEs), smartphones, or wearable devices. The MIMO capacity scaling gain of end users will ultimately be constrained by the number of antennas equipped by UEs. Portable devices, due to their form factor and the cost of implementing dense packaging of antennas, typically have at most 4 or 8 antennas for the frequency range 1 (FR1) spectrum. Additionally, the MIMO multiplexing gain depends on the high-rank channel matrix, which requires sufficient spacing among antenna elements and multipath scattering. Fulfilling these conditions on the UE side may be challenging, limiting support for a high number of spatial streams.

In addition to communication functions, location-based services are becoming increasingly popular in 6G. The 3GPP has been enhancing functionalities that rely on two-way communication between a UE and the 5G Core (5GC) network \cite{Bartoletti-COMSMag21} for more accurate position measurements since Rel-16. However, multiple antennas at the UE side provide an opportunity to enable angle-of-arrival (AoA) estimations. Combining the information from visual-inertial odometry (VIO) systems with the AoA information allows for the positioning process without revealing the UE's position information to the 5GC network, further preserving privacy on the UE side \cite{Shih-IOT23}. However, equipping UEs with large arrays for precise AoA estimation is often infeasible due to form factor constraints.

\begin{figure*}
\centering
\includegraphics[width=5.4in]{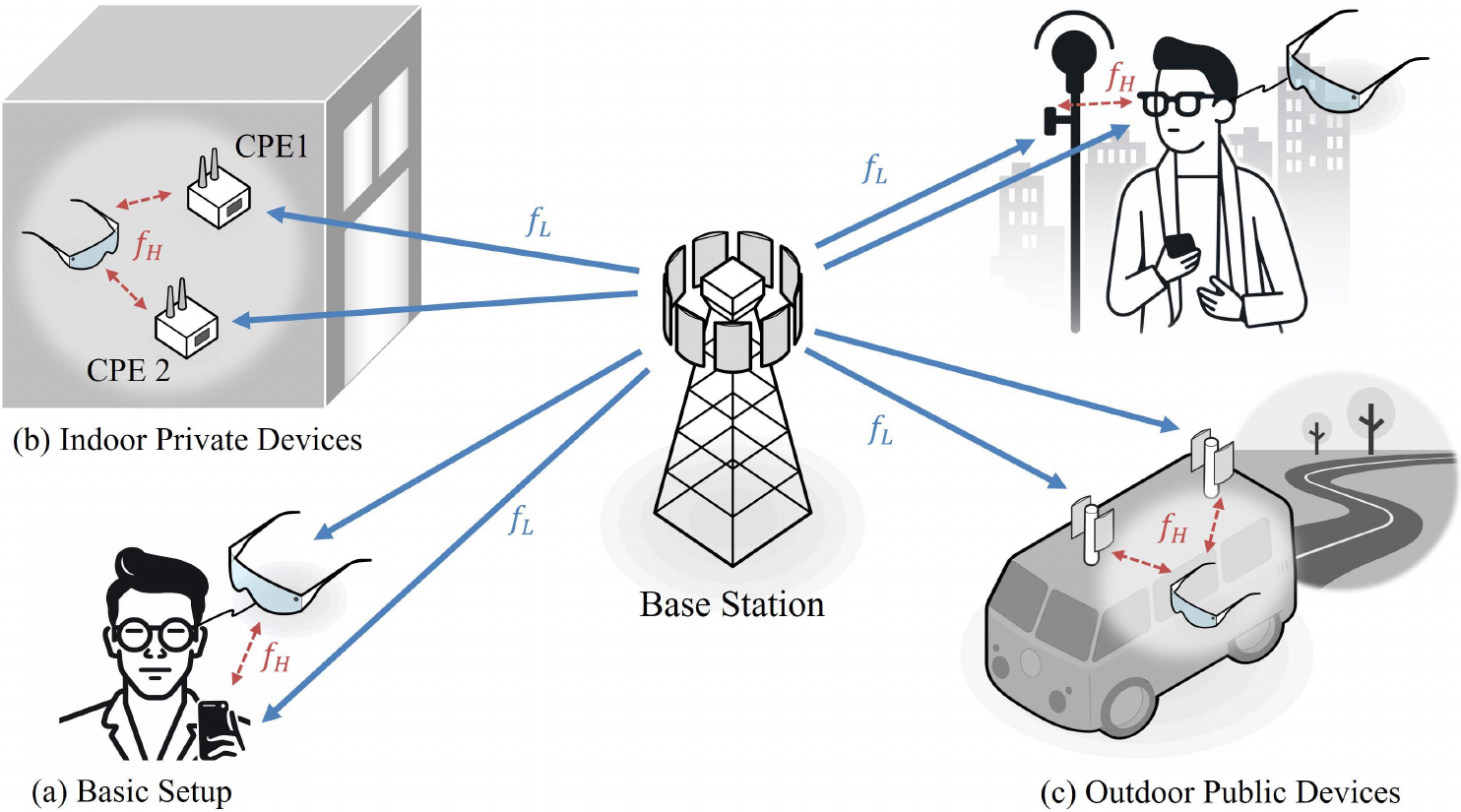}
\caption{Illustration of three exemplary scenarios: (a) basic setup, (b) collaboration with indoor private devices, and (c) collaboration with outdoor public devices.}
\label{fig:PAN_UseCase}
\end{figure*}

To address the limitations of equipping UEs with large arrays, we propose a collaborative MIMO framework called end-user-centric collaborative MIMO (UE-CoMIMO). This framework allows the end user to virtually have numerous antennas, similar to multi-TRP. It has been included in \cite{RWS230111} as one of the new techniques to be discussed in 3GPP Rel-19. As shown in Fig. \ref{fig:PAN_UseCase}, a basic UE-CoMIMO setup involves a smartphone and extended reality (XR) glasses, which are expected to be popular personal area devices. The XR glasses may have two antennas, while the smartphone can be equipped with four antennas. If these devices collaboratively process their received signals, ideally, the effective antenna capacity of the devices can increase to six. From the network's perspective, a BS appears to communicate with an upgraded user that is aggregated by multiple devices.

The proposed UE-CoMIMO framework considers collaboration among devices in a personal area network. To realize the UE-CoMIMO concept, a key enabler is how to connect devices together for collaboration. Wired backhaul connections, commonly used in multi-TRP techniques on the network side, are not ideal for end-user experience and portable/mobile device deployments. Therefore, wireless connections such as Wi-Fi, sidelink, Bluetooth, RF repeater, and L2/L3 relays\footnote{Definitions of L2 and L3 relays are provided in \cite{TR38.836} and \cite{TR23.752}. They are standardized in NR Rel-17. From a protocol stack perspective, the L2 relay is built under packet data convergence protocol, a Layer-2 protocol, while the L3 relay is built under the Layer-3 network layer.} are more feasible to realize the UE-CoMIMO concept. Both L2 and L3 relays need to process protocol-related headers introduced by each layer, decode their received signals to extract data, and reconstruct signals for forwarding. Previous academic studies, such as \cite{Sousa-JWCN13}, have explored the concept of user-side collaboration, focusing on relay selection schemes and theoretical channel capacities under the LTE context, assuming inter-connection among devices relies on a secondary interface like Wi-Fi or Bluetooth. However, conventional L2/L3 relay and Wi-Fi tethering-based approaches introduce latency due to signal decoding and RAT-converting from NR Uu interface to either Wi-Fi or sidelink interface. This latency may make real-time services such as augmented/virtual reality unachievable. Thus, this paper will introduce advanced L1 relay approaches that can aggregate antenna capability among collaborative devices while minimizing latency.

Different personal area devices, such as smartphones, wearable devices, CPEs, or public devices, might have varying capacities in antenna numbers and signal processing, leading to various augmentation modes such as diversity, rank, or localization augmentations. These use cases will be discussed thoroughly in the upcoming sections. The performance improvements of UE-CoMIMO will be quantified by system-level simulations to demonstrate its effectiveness. Lastly, this article will discuss necessary research efforts and directions to realize UE-CoMIMO in future 6G networks.

\section{Scenarios of UE-CoMIMO}
Collaboration among devices, in a user-centric manner, is feasible in both private and public networks comprising multiple wireless-compatible devices. Fig. \ref{fig:PAN_UseCase} illustrates three exemplary scenarios.

\subsubsection{Basic Setup}
UE-CoMIMO should consist of at least two devices. Since most people nowadays have access to a smartphone and some wearable devices, we assume that the basic case of UE-CoMIMO consists of a smartphone and a lightweight device, such as XR glasses, as shown in Fig. \ref{fig:PAN_UseCase}(a). In this scenario, the lightweight devices have limited form factors and processing power, while the smartphone can provide additional performance support to compensate for these limitations, so the performance of the lightweight devices is expected to be enhanced.

\subsubsection{Collaboration with Indoor Private Devices}
The objective of this case is to facilitate collaboration among portable and other private devices within an indoor scenario. For instance, as depicted in Fig. \ref{fig:PAN_UseCase}(b), a smartphone (or XR glasses) can collaborate with CPEs or Wi-Fi access points (APs) installed indoors. Unlike smartphones in the above basic setup, APs or CPEs are usually larger in size and equipped with a power supply. The larger size allows installing more antenna elements and provides spatial isolation among them. Therefore, UE-CoMIMO in this scenario can be more effective than that in the basic setup.

\subsubsection{Collaboration with Outdoor Public Devices}
In outdoor scenarios, the collaboration between infrastructure and portable devices is possible. Infrastructure can include BSs, remote radio heads, or light-pole-mounted devices. For instance, as illustrated in Fig. \ref{fig:PAN_UseCase}(c), XR glasses (or smartphones) can collaborate with light-pole-mounted devices if these devices are available near the user. Device handover is required in this scenario while the UE is moving. Another example is that a user in a vehicle, such as a car, bus, or train, can access services through the XR glasses (or smartphone) by collaborating with antenna modules mounted on or in the vehicle, as shown in Fig. \ref{fig:PAN_UseCase}(c). This application lets the user stay connected to the internet and access various services while moving. Similar to the private devices in the indoor scenario, these public devices also have larger sizes and can provide more effective performance than the basic setup. However, device discovery, authorization, and security issues can present more challenges in these use cases. Additionally, frequent switching between different collaborative devices due to user mobility may introduce latency and complicate handover procedures, which could be subjects for future research.

Overall, the above scenarios demonstrate the versatility of device collaboration and how it can enhance the user's experience in various settings. Notably, all the use cases are initiated in a UE-centric manner. The auxiliary devices are selected by the UE (a smartphone or XR glasses), and these devices may simply forward their received signals to the UE through another frequency band named $f_H$ in Fig. \ref{fig:PAN_UseCase}.

\section{Use Cases of UE-CoMIMO}
\begin{figure*}
\centering
\includegraphics[width=7.2in]{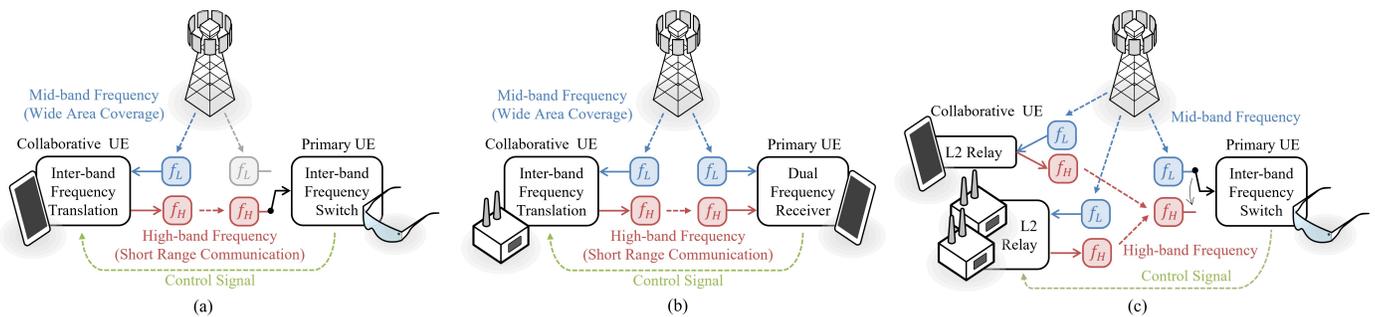}
\caption{Illustration of three types of use cases: (a) diversity augmentation, (b) rank augmentation, and (c) localization augmentation.}
\label{fig:Diversity_Rank_Loc_augmentation}
\end{figure*}

For services that are not latency-sensitive, UE-CoMIMO can rely on conventional L2/L3 relays. However, for services demanding high data rates and low latency, UE-CoMIMO should utilize L1 relays to minimize latency among devices. This work proposes an advanced L1 relay approach to realize UE-CoMIMO among devices with almost zero latency. The approach is achieved by utilizing a frequency-translation amplify-and-forward (AF) repeater, which differs from the conventional same-frequency RF repeater. Specifically, this approach considers two frequency bands, denoted as $f_L$ and $f_H$, which could be frequency bands in FR1 and FR2, respectively. Alternatively, the bands could correspond to a lower mid-band (e.g., 2 GHz) and an upper mid-band (e.g., 7 GHz) in both FR1. It is assumed that the primary UE can receive signals via $f_L$, but may not be able to receive signals from $f_H$ due to high attenuation or blockage at the higher band. Although the $f_H$ band cannot be employed between the BS and UE, it can be utilized for L1 relaying.

This section elaborates on how advanced L1 relays and conventional relays can facilitate device collaboration and addresses three types of use cases: diversity, rank, and localization augmentations.

\subsubsection{Diversity Augmentation}
The first use case corresponds to the basic setup scenario, where UE-CoMIMO aims to provide path diversity gain through device collaboration. In this case, a high-capability device, such as a smartphone, can serve as a proxy to offer a good-quality data path for a low-capability device, like XR glasses. Compared to smartphones, wearable devices typically have fewer capabilities in terms of MIMO processing, the number of transmit/receive antennas (Tx/Rx), and carrier aggregation (CA) capability. Fig. \ref{fig:Diversity_Rank_Loc_augmentation}(a) depicts an example where an intermediate node, such as a 4Rx smartphone, collaborates with a 2Rx wearable device with 1CA capability. Here, the smartphone acts as an advanced AF repeater with superior Rx beamforming capability, i.e., more receive antennas. It enhances signal quality by rejecting interference from unexpected directions and forwards the processed signal to the low-capability wearable device. Furthermore, in such a low-capability device, supporting two frequency bands simultaneously may not be feasible, and thus only $f_H$ band is employed in the primary UE.

Although the path diversity gain discussed above can be achieved by conventional AF RF repeaters, designing a conventional AF RF repeater with adequate physical isolation between its input and output ports is challenging for portable devices with form-factor limitations. Unlike the conventional AF RF repeater design, the proposed L1 relay approach incorporates inter-band frequency translation to overcome full-duplex self-interference issues. This enables the smartphone to perform AF with improved Rx beamforming capability and provide a high-quality data path to the low-capability wearable device, as illustrated in Fig. \ref{fig:Diversity_Rank_Loc_augmentation}(a).

\subsubsection{Rank Augmentation}
The second use case aims to increase the rank of the MIMO channel and can be applied to scenarios involving indoor private devices or outdoor public devices. Fig. \ref{fig:Diversity_Rank_Loc_augmentation}(b) illustrates a scenario where a smartphone and an indoor CPE collaborate to receive a high-rank data signal transmitted from the BS. The smartphone leverages its CA capability to simultaneously receive signals from both the BS and the CPE in two different frequency bands and jointly processes the received signals, effectively simulating two sets of Rx antennas. The BS transmits a signal through $f_L$ band to the primary UE, which now appears to be equipped with two sets of Rx antennas in a single band, resulting in a doubling of the end-to-end MIMO channel rank. By incorporating more CPEs into the network, the number of supported spatial layers can be further increased, bringing it closer to the number of Tx antennas at the BS. To achieve this, the CPEs can forward their received signals through the $f_H$ band.

\subsubsection{Localization Augmentation}
The third use case aims to increase localization precision, which could apply to scenarios with personal portable devices and indoor private devices. For example, determining the location of XR glasses can enhance the user experience. In this case, the XR glasses can find their three-dimensional (3D) location relative to the BS by measuring the distance and 3D angle of the BS from the downlink (DL) signal. However, the two antennas on the XR glasses can only form a linear array, which can measure only 2D AoAs. With UE-CoMIMO, all the antennas of the collaborating devices can form a virtual 3D array, thus enabling the XR glasses to measure 3D AoAs.

The AoA estimation can be performed through synchronization signals (SSs) or DL positioning reference signals (PRSs) in 5G. The SSs are periodically broadcasted (``always-on'') by the BS, while the PRSs are only transmitted on demand. Although AoA estimation through SS blocks does not require additional overhead or bandwidth, the limited bandwidth may lead to insufficient accuracy. If users want to achieve higher accuracy attainable with 5G signals, they can request wide-band reference signals such as PRSs to use the entire system bandwidth or cognitively utilize other reference signals \cite{Neinavaie-22JSTSP}.

Since the location-based service is insensitive to symbol-level latency, necessary information for estimating 3D AoA can be collected through conventional L2/L3 relays. As precise inter-device timing synchronization may not be achieved, a non-coherent AoA estimation scheme (e.g., \cite{Rath-ACCESS20}) that combines estimated channel responses from multiple collaborative devices can be used. Specifically, the XR glasses receive processed information (e.g., channel responses estimated in $f_L$ band) from the collaborative smartphone or CPEs through $f_H$ band, as illustrated in Fig. \ref{fig:Diversity_Rank_Loc_augmentation}(c). With this processed information from the collaborative devices, the XR glasses can estimate 3D AoA jointly.

\section{Performance Evaluations}

\subsection{UE-CoMIMO Aided by Advanced L1 Relay}

In this subsection, we conduct system-level simulations following the 3GPP evaluation methodology to illustrate the potential performance benefits of UE-CoMIMO aided by the advanced L1 relay. The simulations will consider 3GPP dense urban scenarios as specified in TR 38.913 \cite{TR38.913}, with an inter-site distance of 200 m. All UEs are indoor users, who may experience significant path loss for DL reception or uplink (UL) transmission. For advanced UEs that support the schemes depicted in Fig. \ref{fig:Diversity_Rank_Loc_augmentation}(a) or Fig. \ref{fig:Diversity_Rank_Loc_augmentation}(b), each of them will be paired with a collaborative UE supporting frequency-translation located 1 m away.
The simulations in this work assume the use of a 5G NR orthogonal frequency-division multiplexing (OFDM) waveform. The BS adopts a modulation order and code rate based on either channel state information reported by the primary UEs for DL or measurements at the BS by receiving sounding reference signal transmitted by the primary UEs for UL. Other detailed parameters are listed in Table \ref{tab:SimScenario}.

\begin{table}[!t]
\caption{Simulation Scenario}\label{tab:SimScenario}
\centering
\begin{tabular}{ | c | c|  }
  \hline
  \multirow{4}{12em}{\centering Scenario}   & Dense Urban (Hex. Grid); \\
                                &  3 BSs per site;   \\
                                & site-to-site distance = 200m; \\
                                & indoor UE; 10 UEs/cell  \\
  \hline
  \multirow{2}{12em}{\centering Traffic Type}   & Full-buffer traffic or \\
                                &   FTP traffic   \\
  \hline
  Carrier Frequency & $(f_L,f_H) = (2\,{\rm GHz}, 6\,{\rm GHz})$ \\
  \hline
  System bandwidth	& 10 MHz \\
  \hline
  Subcarrier spacing &	30 kHz \\
  \hline
  Number of antenna Rx  & \multirow{2}{12em}{\centering 32 ports} \\
  elements per BS       &  \\
  \hline
  Number of antenna Tx/Rx  & 2T4R (DL) \\
  elements per UE       &  2T2R (UL) \\
  \hline
  Distance between & \multirow{3}{12em}{\centering 1 m} \\
  collaborative UE &    \\
  and primary  UE  &    \\
  \hline
  UE max Tx power  & 23 dBm \\
  \hline
  Scheduling  & Proportional fair \\
  \hline
  Receiver type  & MMSE-IRC \\
  \hline
  UL channel sounding  & 5 slots period \\
  \hline
  DL precoding  & NR type-II codebook \\
  \hline
  UL precoding  & SVD based \\
  \hline
\end{tabular}
\end{table}

\subsubsection{Diversity Augmentation}
\begin{figure}
\centering
\includegraphics[width=3.3in]{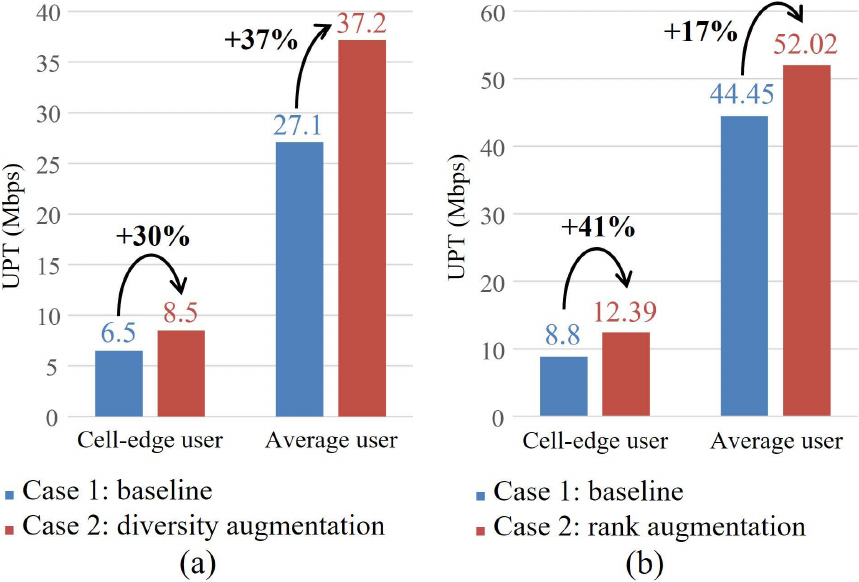}
\caption{User throughput comparison between Case 1 and Case 2 for a medium cell load (RU = 40\%). (a) With and without diversity augmentation (b) With and without rank augmentation}
\label{fig:Performance_Rank}
\end{figure}

We compare the DL user-perceived throughput (UPT) for two scenarios under a full-buffer traffic model and FTP traffic model 3 defined in 3GPP with 10 UEs/cell/band:

\noindent {\bf Case 1:} This is the baseline scenario where all UEs receive DL signals only in $f_L$.

\noindent {\bf Case 2:} All advanced UEs are capable of performing path selection shown in Fig. \ref{fig:Diversity_Rank_Loc_augmentation}(a) to receive a signal in either $f_L$ or $f_H$. Meanwhile, to check whether the BS's transmission in $f_H$ has an impact on the advanced UEs, we assume all BSs serve other legacy UEs in $f_H$.

The simulation results show that, under full-buffer traffic model, the proposed scheme provides additional gains of +13.1\% and +18.3\% in cell-edge and average UPT, respectively, for the advanced UEs. The performance degradation of the legacy UEs in $f_H$ is mild, with -3.7\% and -2.6\% in cell-edge and average UPT, respectively. In another setting under 3GPP FTP traffic model with medium cell load (where resource utilization rate is about 40\%), the proposed scheme provides additional gains of +30\% and +37\% in cell-edge and average UPT, respectively, as shown in Fig. \ref{fig:Performance_Rank}(a).

\subsubsection{Rank Augmentation}
We compare the UL UPT for two cases under the FTP traffic model:

\noindent {\bf Case 1:} Legacy NR 2CA scheme with direct transmission from UE to BS in both $f_L$ and $f_H$.

\noindent {\bf Case 2:} UEs additionally support the proposed device collaboration scheme shown in Fig. \ref{fig:Diversity_Rank_Loc_augmentation}(b) for a medium cell load (resource utilization rate is about 40\%)

For this evaluation, we assume that all primary UEs can support 2CA and device collaboration. The scheme used for each UE is semi-statically determined by the network according to the signal quality in $f_L$ and $f_H$. For example, for a UE that is out-of-coverage in the high band $f_H$, since the high band is no longer suitable for direct UL transmission of the UE, the network may enable device collaboration for rank-boosting in $f_L$ with negligible impact on other co-channel UL links in $f_H$.

As shown in Fig. \ref{fig:Performance_Rank}(b), the proposed scheme provides a +41\% and +17\% additional gain in cell-edge and average UPT, respectively, for the medium-load case.

\subsection{UE-CoMIMO Aided by L2/L3 Relay}

\begin{figure}
\centering
\includegraphics[width=3.5in]{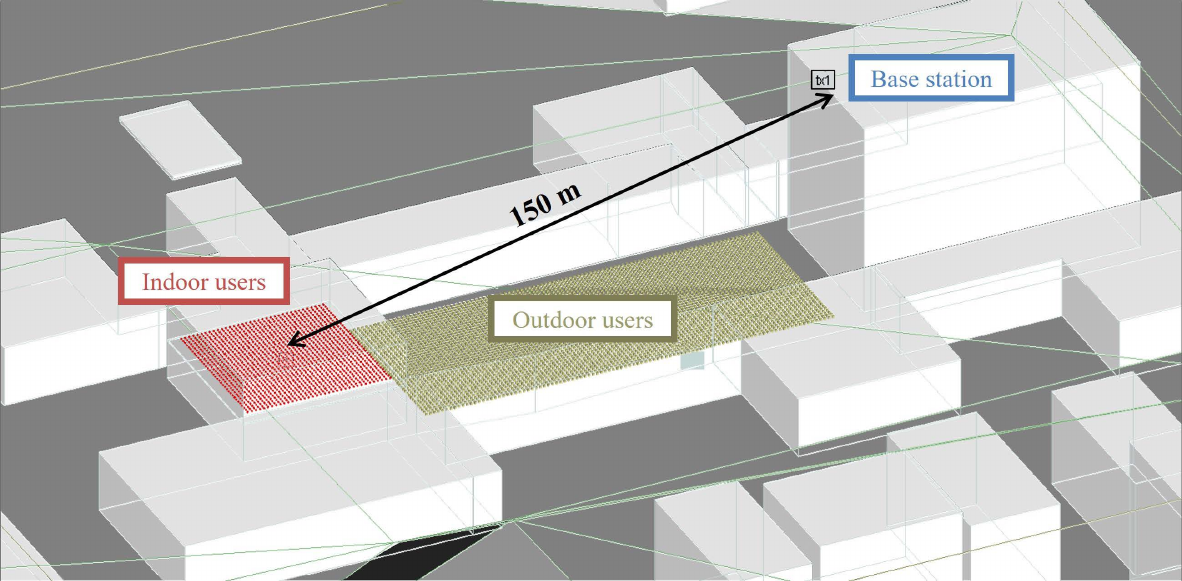}
\caption{Position of the base station, the outdoor users, and the indoor users in the building of the Wireless Insite simulation.}
\label{fig:Wireless Insite}
\end{figure}

In this subsection, we conduct simulations to illustrate the performance benefits of localization augmentation achieved through the conventional L2/L3 relay. The testing scenario is depicted in Fig. \ref{fig:Wireless Insite}, where radio wave propagation is generated using a suite of ray-tracing models, Wireless InSite$\circledR$. We set the receive points one meter away from the ground in the outdoor environment as the ``outdoor users'' and set the receive points one meter away from the floor in the middle of the building as the ``indoor users.'' The AoA estimation is performed using the SSs transmitted from only a single BS.

We compare the estimation performances for the following three cases:

\noindent {\bf Case 1:} This is the baseline scenario where 3D AoA is estimated by 2Rx XR glasses.

\noindent {\bf Case 2:} This is a scenario where 3D AoA is estimated by collaborating 2Rx XR glasses as the primary UE and a 2Rx smartphone as the collaborative UE, as shown in Fig. \ref{fig:PAN_UseCase}(a).

\noindent {\bf Case 3:} This scenario assumes that portable devices are in the building and can collaborate with CPE-like private devices. Besides 2Rx XR glasses and one 2Rx smartphone, two additional 4Rx CPEs are used as the collaborative UEs, as illustrated in Fig. \ref{fig:PAN_UseCase}(b).

\begin{figure}
\centering
\includegraphics[width=3.5in]{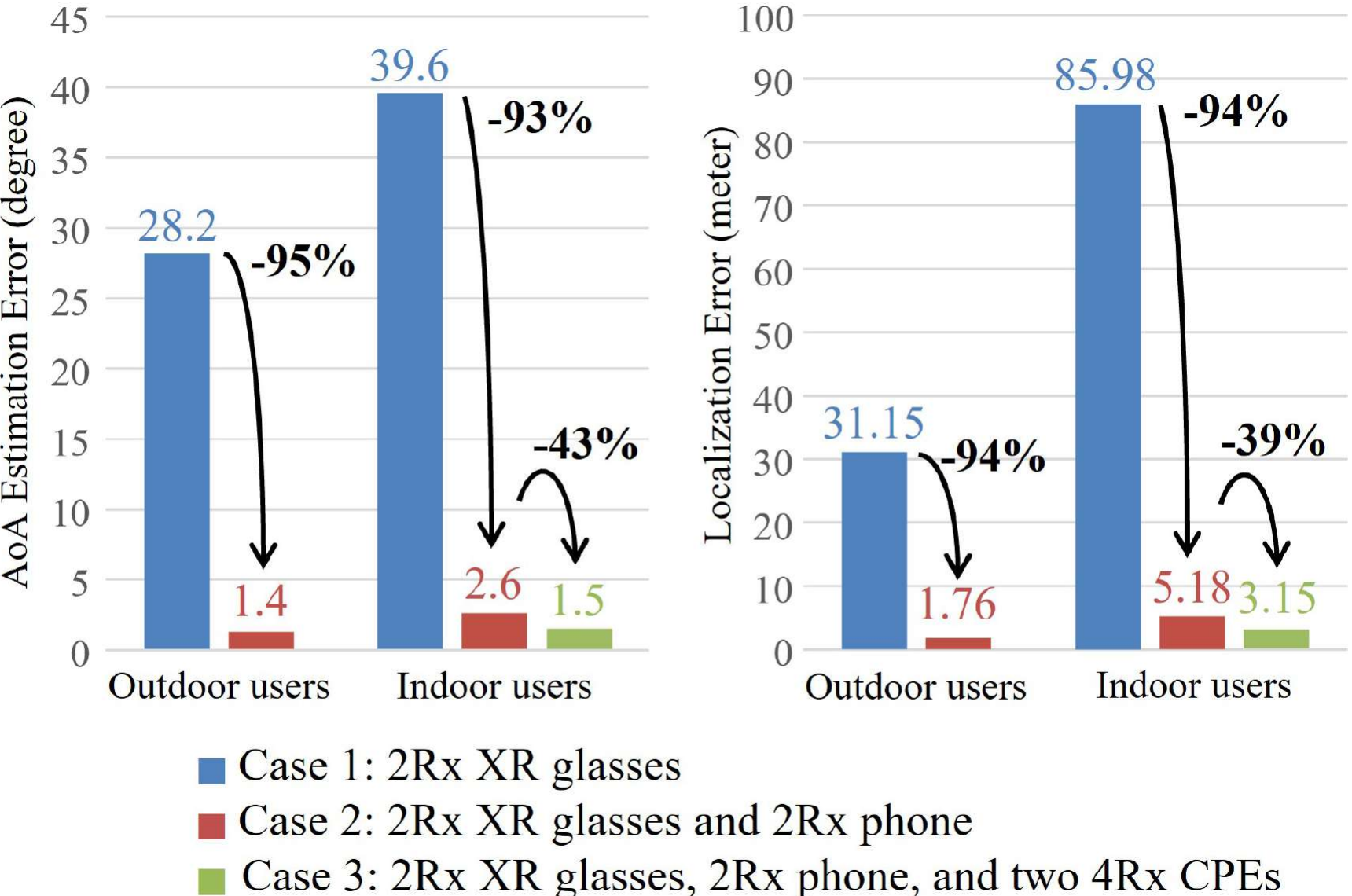}
\caption{3D AoA estimation and localization comparison between Case 1, Case 2, and Case 3.}
\label{fig:Localization_Rank}
\end{figure}

Our results in Fig. \ref{fig:Localization_Rank} show that the baseline scenario (i.e., 2Rx XR glasses only) cannot conduct 3D AoA estimation based on SSs from a single BS alone, resulting in dreadful performance. However, by aggregating antennas from nearby portable devices, the estimation becomes feasible. Case 2 exhibits a 95\% and 93\% reduction in AoA estimation error for outdoor and indoor users, respectively. Moreover, when portable devices in the building collaborate with more private devices, compared to Case 2, Case 3 shows a 43\% further reduction of AoA estimation error, and the localization accuracy is improved by 2.03 meters.

\section{Research Directions}
In this section, we will discuss critical research efforts for integrating UE-CoMIMO into 6G and related research areas.

\subsubsection{Additional Spectrum and Standardization}
One key characteristic of UE-CoMIMO is the need for additional spectrum for local links among collaborative devices. Notably, the proposed UE-CoMIMO schemes differ from direct communications that use double spectrum because the BS does not use double spectrum. However, if the licensed band is reused directly, there may be concerns about whether the assisted devices cause interference to legacy operations in the same band. Fortunately, the path loss of the channel between any two devices in proximity is typically small, so the transmission power of the L1 relay node is usually not large (e.g., less than 14 dBm) and is unlikely to cause serious problems to the network. Alternatively, the network may select an unused or lowly utilized band to minimize the negative impact. For instance, if the end user is in a coverage hole of a particular band, which could be a high-frequency band, the band could be a good candidate for the links among collaborative devices.

Understanding the spectrum requirements and exploring the possibilities for efficient spectrum utilization will be crucial for UE-CoMIMO implementation in 6G networks. Research should evaluate benefits and drawbacks of various frequency bands for local links, investigate interference management techniques, and study dynamic spectrum allocation strategies for resource efficiency and seamless integration with existing and emerging wireless technologies.

Additionally, to popularize the UE-CoMIMO technique, practical aspects including hardware requirements, control signaling supports, and interoperability across devices from different vendors are very important. Some standardization works are needed in standard bodies to realize the proposed techniques. In \cite{RWS230111}, potential standardization efforts to realize the proposed techniques were examined.

\subsubsection{Inter-FR translation}
High-frequency bands, such as FR2 bands in 5G, may be considered excellent candidate spectrums for local links among collaborative devices due to the attenuation nature of high-frequency signals. They are beneficial to avoid mutual interference among each cluster of collaborative devices. However, when frequency translation is needed across two different FRs, subcarrier spacing (SCS) configurations in two FRs are typically different (e.g., 30 kHz in FR1 and 120 kHz in FR2), making the same-SCS frequency translation repeater unsuitable for inter-FR translation.

One solution is to keep signals in the local links following the same SCS as that in FR1. However, this approach may not be optimal for FR2 transmission and reception, where a larger SCS is adopted to mitigate phase noise effects. Another solution is to perform frequency translation in the frequency domain using the fast Fourier transform (FFT). This involves extracting the baseband signal in the frequency domain from the received signal of the L1 relaying node in FR1 and then amplifying and forwarding the frequency-domain baseband signal by an RF signal to be transmitted in FR2, where a larger SCS is adopted to generate the retransmitted OFDM signal. However, this solution introduces additional latency due to FFT and inverse FFT conversions. To achieve an efficient and low-latency inter-FR translation for UE-CoMIMO, future research should focus on addressing the challenges associated with different SCS configurations and the impact of phase noise.

\subsubsection{Virtual Array}
The AoA can be estimated from the received RF signals based on the antenna array response, which depends on the position of the antennas on the array. To collaborate multiple UEs for 3D AoA estimation, the primary UE requires information about the relative distance, bearing, and antennas' position of the collaborative UE to form the virtual 3D array. The distance and bearing of the collaborative UE can be obtained from the VIO on the primary UE, as the VIO incorporates a camera and an inertial measurement unit. However, for the XR glasses to form a complete 3D array system, it is necessary to obtain the position of the antennas inside the collaborative UE. This process can be easily achieved when the devices belong to the same company, but obtaining this information may be more challenging if the devices are from different companies.

One solution could be to standardize devices willing to provide UE-CoMIMO services with other companies' devices, enabling them to share this information through sidelink or Wi-Fi. Another solution is that each device estimates AoA itself without doing signal-level combining, then reports its estimation result for further combining, which degenerates to a conventional cooperative localization method \cite{Wymeersch-09IEEEProc}. For localization augmentation, the former solution is more suitable, where personal area devices can form a large virtual 3D array. Future research should focus on integrating local location information from VIO systems and global location information embedded in cellular signals to improve location precision and robustness.

\subsubsection{Localization Scheme}
When collaborating devices are close to each other, they can estimate a common angle together based on a far-field signal model applied to received RSs. However, the angles between the BS and each device vary when the devices are farther apart, which implies that assuming a common angle would cause errors. In this case, the devices can form an extra-large antenna array, and the DL signals received by the devices should be near-field signals, allowing the relative location between the BS and the devices to be directly estimated. Although the location can be obtained without separately estimating distance and angle based on the near-field model, the calculation complexity of the direct localization method is significantly high. Developing localization methods with limited calculation complexity is a potential future research area for collaborating devices.

In this study, we employ a non-coherent AoA estimation scheme without strict inter-device timing synchronization. A coherent scheme can estimate AoA more accurately, as all collaborating devices are synchronized to a common reference clock. This can be achieved by using a smartphone as a central device to transmit bursts of modulated radio signals for timing acquisition \cite{Koelemeij-22Nature}.

\subsubsection{Integration with Reconfigurable Intelligent Surfaces}
This study uses L1 relay approaches to aggregate antenna capability among collaborative devices. However, there are other potential solutions, such as in-band AF RF repeaters and reconfigurable intelligent surfaces (RISs), which also introduce negligible delay. For active devices, an in-band AF repeater must solve the problem of self-interference coupled from transmission antennas to reception antennas, which is a challenge to realize on a portable device due to its limited form factor. For passive devices, an RIS also requires a large panel size to provide sufficient link supplement. Both technologies offer a promising way to improve channel conditions through rank or diversity supplementation without requiring additional spectrum. However, they cannot provide additional virtual receiver numbers like the proposed L1 relay approaches, and thus the MIMO capacity scaling gain is still limited by the number of antennas of the primary UE. Using RISs could help mitigate the impact of the surrounding environment and reflections, while UE-CoMIMO could provide additional virtual receivers to further enhance the system's performance. Therefore, integrating UE-CoMIMO with RIS technology could be a potential research direction for future work.

\subsubsection{Other Potential Augmentation}
In addition to the diversity, rank, and localization augmentations discussed in this study, the increasing antenna number at the UE side brings possibilities for other types of augmentations. One such augmentation is sensing augmentation. For instance, in the context of 5G NR, a user can instruct the XR glasses to perform UL beam sweeping while using a smartphone to measure the reflected signal, effectively creating a MIMO radar. Changing the relative positions of the devices through different gestures can improve the resolution of sensing. The observed surrounding environment through the active sensing provides situational awareness, which can be used to enhance communication and localization performances \cite{Fan-JSAC22,Yang-JSAC22}.

\section{Conclusion}
Rather than following the traditional path of MIMO evolution from the network side, this article has explored an alternative approach, focusing on the evolution of MIMO from the UE side and introducing the innovative concept of UE-CoMIMO. This method leverages L1 relay-based techniques in conjunction with traditional L2/L3 relays, allowing for the flexible creation of an expansive MIMO system at the user-end in a decentralized manner. We have explored its potential through three types of augmentation: diversity, rank, and localization augmentations, and demonstrated their performance gains through system-level simulations. The applications of UE-CoMIMO are wide-ranging, including improving the cell-edge user experience, increasing network capacity, and enhancing location and sensing precision. In conclusion, while network-based MIMO evolution technologies, such as network densification and network-side cooperation, effectively enhance user experience, it is essential to recognize that UE-CoMIMO can complement and integrate seamlessly with these methods, driving towards optimal outcomes.



\begin{thebibliography}{10}
\providecommand{\url}[1]{#1}
\csname url@samestyle\endcsname
\providecommand{\newblock}{\relax}
\providecommand{\bibinfo}[2]{#2}
\providecommand{\BIBentrySTDinterwordspacing}{\spaceskip=0pt\relax}
\providecommand{\BIBentryALTinterwordstretchfactor}{4}
\providecommand{\BIBentryALTinterwordspacing}{\spaceskip=\fontdimen2\font plus
\BIBentryALTinterwordstretchfactor\fontdimen3\font minus
  \fontdimen4\font\relax}
\providecommand{\BIBforeignlanguage}[2]{{%
\expandafter\ifx\csname l@#1\endcsname\relax
\typeout{** WARNING: IEEEtran.bst: No hyphenation pattern has been}%
\typeout{** loaded for the language `#1'. Using the pattern for}%
\typeout{** the default language instead.}%
\else
\language=\csname l@#1\endcsname
\fi
#2}}
\providecommand{\BIBdecl}{\relax}
\BIBdecl

\bibitem{Jin-23JSAC}
H.~Jin, K.~Liu, M.~Zhang, L.~Zhang, G.~Lee, E.~N. Farag, D.~Zhu,
  E.~Onggosanusi, M.~Shafi, and H.~Tataria, ``Massive {MIMO} evolution toward
  {3GPP} {Release} 18,'' \emph{IEEE J. Sel. Areas Commun.}, vol.~41, no.~6, pp.
  1635--1654, 2023.

\bibitem{Jeon-21COMMag}
J.~Jeon, G.~Lee, A.~A. Ibrahim, J.~Yuan, G.~Xu, J.~Cho, E.~Onggosanusi, Y.~Kim,
  J.~Lee, and J.~C. Zhang, ``{MIMO} evolution toward {6G}: Modular massive
  {MIMO} in low-frequency bands,'' \emph{IEEE Commun. Mag.}, vol.~59, no.~11,
  pp. 52--58, 2021.

\bibitem{Bartoletti-COMSMag21}
S.~Bartoletti, H.~Wymeersch, T.~Mach, O.~Brunnegård, D.~Giustiniano,
  P.~Hammarberg, M.~F. Keskin, J.~O. Lacruz, S.~M. Razavi, J.~Rönnblom,
  F.~Tufvesson, J.~Widmer, and N.~B. Melazzi, ``Positioning and sensing for
  vehicular safety applications in {5G} and beyond,'' \emph{IEEE Comm. Mag.},
  vol.~59, no.~11, pp. 15--21, 2021.

\bibitem{Shih-IOT23}
W.-T. Shih, C.-K. Wen, S.-H. Tsai, R.~Liu, and C.~Yuen, ``{EasyAPPos}:
  Positioning {Wi-Fi} access points by using a mobile phone,'' \emph{IEEE
  Internet Things J.}, vol.~10, no.~15, pp. 13\,385--13\,400, 2023.

\bibitem{RWS230111}
RWS-230111, ``Device collaborative {T}x and {R}x,'' MediaTek Inc., 3GPP RAN
  Rel-19 workshop, Taipei, TW, Jun. 2023.

\bibitem{TR38.836}
{3GPP TR 38.836}, ``Study on {NR} sidelink relay,'' V17.0.0, Mar. 2021.

\bibitem{TR23.752}
{3GPP TR 23.752}, ``{Study on system enhancement for Proximity based Services
  (ProSe) in the 5G System (5GS)},'' V17.0.0, Mar. 2021.

\bibitem{Sousa-JWCN13}
I.~Sousa, M.~P. Queluz, and A.~Rodrigues, ``A smart scheme for relay selection
  in cooperative wireless communication systems,'' \emph{J Wireless Com
  Network}, vol. 2013, no.~1, 2013.

\bibitem{Neinavaie-22JSTSP}
M.~Neinavaie, J.~Khalife, and Z.~M. Kassas, ``Cognitive opportunistic
  navigation in private networks with {5G} signals and beyond,'' \emph{IEEE J.
  Sel. Topics Signal Process}, vol.~16, no.~1, pp. 129--143, 2022.

\bibitem{Rath-ACCESS20}
M.~Rath, J.~Kulmer, E.~Leitinger, and K.~Witrisal, ``Single-anchor positioning:
  Multipath processing with non-coherent directional measurements,'' \emph{IEEE
  Access}, vol.~8, pp. 88\,115--88\,132, 2020.

\bibitem{TR38.913}
{3GPP TR 38.913}, ``Study on scenarios and requirements for next generation
  access technologies,'' V14.1.0, Dec. 2016.

\bibitem{Wymeersch-09IEEEProc}
H.~Wymeersch, J.~Lien, and M.~Z. Win, ``Cooperative localization in wireless
  networks,'' \emph{Proc. IEEE}, vol.~97, no.~2, pp. 427--450, 2009.

\bibitem{Koelemeij-22Nature}
{J. C. J. Koelemeij, H. Dun, C. E. V. Diouf, E. F. Dierikx, G. J. M. Janssen,
  and C. C. J. M. Tiberius}, ``A hybrid optical–wireless network for
  decimetre-level terrestrial positioning,'' \emph{Nature}, vol. 611, no. 7936,
  pp. 473--478, Nov. 2022.

\bibitem{Fan-JSAC22}
F.~Liu, Y.~Cui, C.~Masouros, J.~Xu, T.~X. Han, Y.~C. Eldar, and S.~Buzzi,
  ``Integrated sensing and communications: Toward dual-functional wireless
  networks for {6G} and beyond,'' \emph{IEEE J. Sel. Areas Commun.}, vol.~40,
  no.~6, pp. 1728--1767, 2022.

\bibitem{Yang-JSAC22}
J.~Yang, C.-K. Wen, and S.~Jin, ``Hybrid active and passive sensing for {SLAM}
  in wireless communication systems,'' \emph{IEEE J. Sel. Areas Commun.},
  vol.~40, no.~7, pp. 2146--2163, 2022.

\end{thebibliography}
\end{document}